\documentclass{article}

\usepackage{PRIMEarxiv}

\usepackage{hyperref}       
\usepackage{url}            
\usepackage{cite}
\usepackage{amsmath,amssymb,amsfonts}
\usepackage{graphicx}
\usepackage{color}

\title{Generalised Impedance Model of Wireless Links Assisted by Reconfigurable Intelligent Surfaces
}

\author{
  Keisuke Konno,  Qiang Chen \\
  Department of Communications Engineering \\
  Graduate School of Engineering \\
  Tohoku University \\
  Japan\\
  \texttt{\{konno,qiang.chen.a5\}@ecei.tohoku.ac.jp} \\
   \And
  Sergio Terranova, Gabriele Gradoni \\
  School of Mathematical Sciences\\
  University of Nottingham \\
  United Kingdom\\
  \texttt{\{sergio.terranova,gabriele.gradoni\}@nottingham.ac.uk} \\
}

\begin{document}
\maketitle

\begin{abstract}
We devise an end-to-end communication channel model that describes the performance of RIS-assisted MIMO wireless links. 
The model borrows the impedance (interaction) matrix formalism from the Method of Moments and provides a physics-based 
communication model.
In configurations where the transmit and receive antenna arrays are distant from the RIS beyond a wavelength, 
a reduced model provides accurate results for arbitrary RIS unit cell geometry.
Importantly, the simplified model configures as a cascaded channel transfer matrix whose mathematical structure 
is compliant with widely accepted, but less accurate, system level RIS models. 
A numerical validation of the communication model is presented for the design of binary RIS structures 
with scatterers of canonical geometry. 
Attained results are consistent with path-loss models: For obstructed line-of-sight between transmitter and receiver, 
the channel capacity of the (optimised) RIS-assisted link scales as $R^{-2}$, 
with $R$ RIS-receiver distance at fixed transmitter position.
Our results shows that the applicability of communication models based on mutual impedance matrices is not restricted to canonical minimum 
scattering RIS unit cells.
\end{abstract}
\vspace{5mm}
    {\small \bf
    \textbf{NOTE:} This work has been submitted to the IEEE TAP for possible publication.\\
    Copyright may be transferred without notice, after which this version may no longer be accessible.
    }

\bigskip

\noindent 

\section{Introduction}
Reconfigurable intelligent surfaces (RIS) provides a viable technology for molding the flow of electromagnetic wave energy within 
arbitrary propagation environments \cite{ECS1}.
Wave control capabilities demonstrated by the RIS include anomalous reflection, beam-forming, focusing, and defocusing \cite{MdR1}.
Their operation has been studied in free-space, both near-field and far-field \cite{ZSG1,BAD1}, as well as in indoor and outdoor environments \cite{Pei1}.
The modelling of RIS-assisted wireless links have became a topic of paramount importance \cite{GMdR1}, and several authors have proposed 
physics-based communications models \cite{MdR2}. 
The scientific literature classifies first-principle electromagnetics model of the RIS in two families: Continuous and Discrete models \cite{MdR3}.
Continuous models treat meta-surfaces as homogeneous sheets of currents whole effective impedance and admittance are matched by design to perform 
anomalous reflection.
Discrete models treat individual pixels: In the first models inspired from antenna arrays, the effective scattered field results from the superposition of 
waves re-radiated upon specular reflection of individual RIS unit cells. 
However, it has been shown that effective models based on re-radiated fields: i) do not correctly capture the mutual coupling between unit cells within 
finite size RIS structures \cite{Marco,Robin}; ii) do rely on local periodicity which provides an approximate behaviour of the field re-radiated by finite-size macro cells \cite{Vittorio}; 
iii) do not accurately capture the effect of loading circuitry on the re-radiated field, as well as parasitic effects within the unit cell \cite{Hum}.
More recently, discrete models of the RIS have been put forward using both impedance \cite{WRY1} and scattering \cite{SCM1} matrix formalisms. 
Those models are capable of treating a finite-size RIS, capturing joint amplitude and phase variability of the RIS unit cell response, 
and providing an end-to-end (E2E) representation of the wireless link, i.e., include the interaction of the RIS with the transmit and receive arrays 
associated to the base-station (BS) to user-equipment (UE) respectively.
A specific impedance-based channel model has been formulated for arrays of dipoles \cite{GD1}, and exploited to analyse the benefit of 
the knowledge of mutual coupling between dipoles in channel gain optimisation \cite{QMdR1}.
While the model works in both near- and far-field, the RIS unit cell has been assumed to be a minimal scattering linear antenna, for which the surface current density 
is known in closed form and the array active impedance can be computed efficiently including mutual coupling \cite{MdRGC1}. 
The structural scattering of the open-circuit \cite{MH1} baseline RIS structure needs to be considered and included in impedance-based models.
We extend the model in \cite{GD1} to arbitrary arrays of scatterers adopted for RIS using the Method-of-Moment (MoM) formalism.
The MoM code that we use has been recently developed in \cite{Arbitrary-load1}, which has used the methodology originally proposed in the theory of loaded scatterers \cite{MH1},  
where port segments have been separated from passive segments in the MoM discretisation.
A MoM study has been recently proposed for the path loss of a RIS-assisted SISO system, where freestanding loaded loops have been assumed 
as RIS unit cells \cite{SA1}.
In our MoM model, the ground plane has been included and the formulation is amenable to accounting for single- as well as multi-layer substrates 
by using standard methodologies \cite{LMGF4}.
Furthermore, our model provides an end-to-end communication model that can be conveniently used for optimisation and performance analysis of wireless links, 
and it naturally extends the model in \cite{GD1} beyond the canonical minimum scattering (CMS) approximation. 
This extension has the benefit of making important work on CMS structures practical: This work has been achieved in the literature lately, also considering 
wide-band response of linear antennas \cite{ABMN1}, and including e.g., multi-path fading generated by clusters of dipoles in the impedance \cite{MPSGDR1} 
and the Green function \cite{FSASIdH1} formalisms. 
Previous studies have discussed the inclusion of circuital elements within periodic meta-surfaces with sub-wavelength unit cells in MoM \cite{HVSC1}, 
There, a simple development of the MoM interaction matrix allowed the computation of the scattered field for arbitrary distribution of discrete loads.
This boils down to add the impedance matrix of the equivalent circuit of the tuneable elements to the MoM interaction impedance matrix of port basis segments. 
In RIS design, it is worth noticing that the inner coupling between loading circuitry and planar metallic structure might generate parasitic effects, 
which are usually captured by a more complicated (and structure dependent) equivalent circuit \cite{HOD1}.
In the far-field of a single RIS unit cell, our model reduces to a cascaded formula that has the same mathematical structure of system levels 
MIMO communication models \cite{GD1}.
However, the impedance matrix are augmented with the self- and mutual- interactions involving scattering segments besides port segments.
Importantly, while system level models are not constructed in physics-based grounds and are prone to efficient optimisation, 
our electromagnetic compliant model preserves accuracy for arbitrary RIS unit cell geometry.
A numerical analysis of the the simplified communication model has been carried out for two canonical unit cell geometries, 
and validated by the (related) exact MoM code. 
The performance of a compact RIS-assisted MIMO link are also assessed in terms of channel capacity.

The paper is organised as follows, in Sec. II we apply the method-of-moment formulation to the communication system 
including the RIS, transmit and receive arrays. 
We partition basis functions into port and passive segments.
We then devise an end-to-end communication model, and devise a reduced cascaded model.
In Sec. III we design two RIS structures based on loaded scatterers, and we assess the model by numerical validations 
of the channel capacity at variable receiver distance and direction. 
Finally, in Sec. IV we draw a conclusion and future perspectives.

\section{Formulation}
\subsection{Full-wave model}
\begin{figure}[t]
\begin{center}
\includegraphics[width=8.5cm]{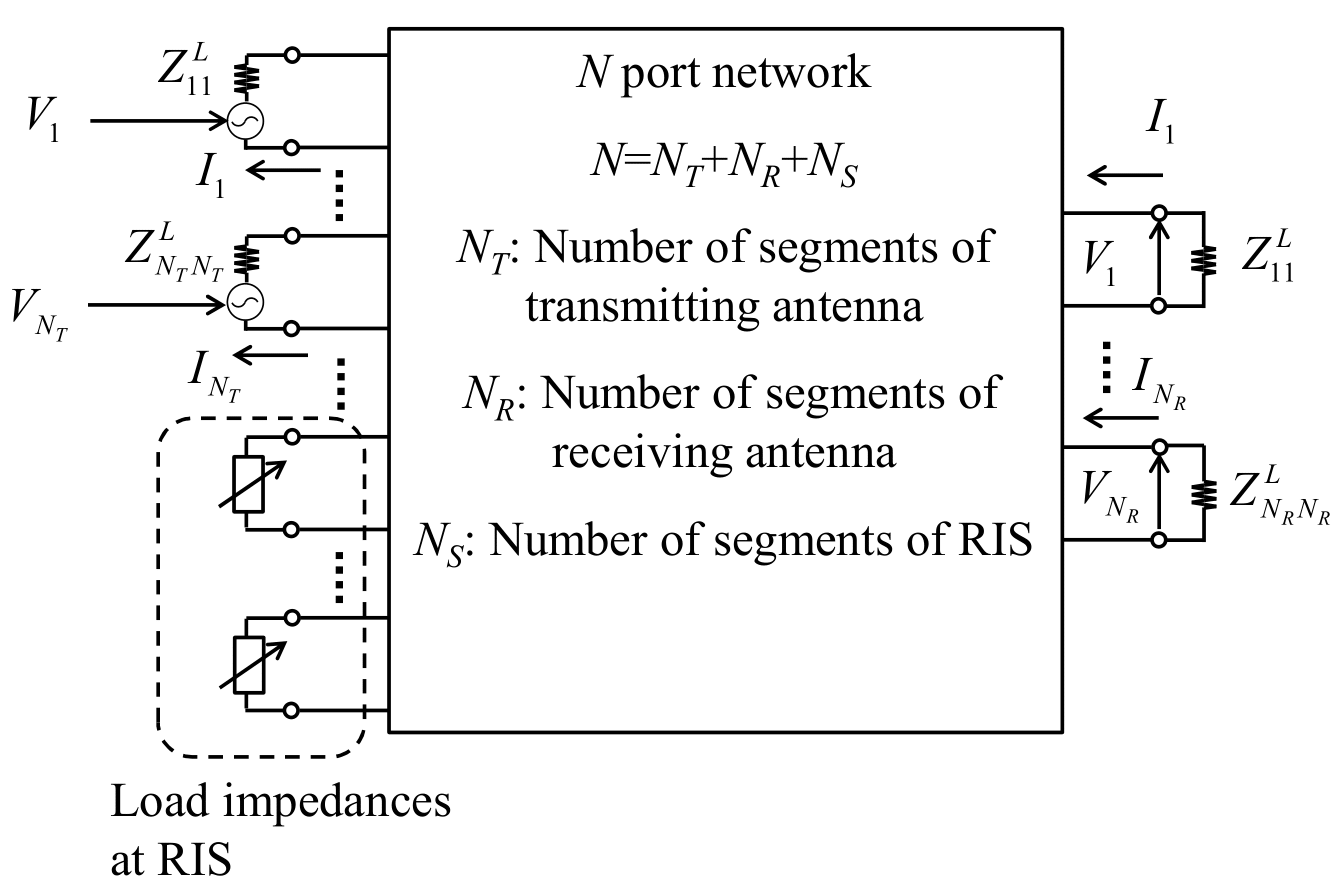}\\
\caption{Equivalent circuit of RIS including transmitting/receiving antennas. }
\label{fig:Network}
\end{center}
\end{figure}
Here, an equivalent circuit of RIS including transmitting/receiving antennas is modeled via method of moments (MoM) as shown in Fig. \ref{fig:Network}. 
The equivalent circuit is expressed as a matrix equation as follows. 
\begin{equation}
({\mathbf{Z}}_{N \times N}  + {\mathbf{Z}}_{N \times N}^L ){\mathbf{I}}_N  = {\mathbf{V}}_N 
\label{eq:Matrix-1}
\end{equation}
Here, ${\mathbf{Z}}_{N \times N}$ is $N \times N$ impedance matrix, ${\mathbf{Z}}_{N \times N}^L$ is $N \times N$ load impedance matrix, ${\mathbf{I}}_N$ is $N$ dimensional current vector, and ${\mathbf{V}}_N$ is $N$ dimensional voltage vector, and $N$ is the total number of segments (i.e. number of unknowns).  

Equation (\ref{eq:Matrix-1}) can be partitioned using block matrices/vectors as follows.  
\begin{align}
  \left[ {\begin{array}{*{20}c}
   {\bf{V}}_{N_T}   \\
   0  \\
   0  \\
 \end{array} } \right] &= \left[ {\begin{array}{*{20}c}
   {{\bf{Z}}_{N_T  \times N_T }  + {\bf{Z}}_{N_T  \times N_T }^L } & {{\bf{Z}}_{N_T  \times N_R } } \\
   {{\bf{Z}}_{N_R  \times N_T } } & {{\bf{Z}}_{N_R  \times N_R }  + {\bf{Z}}_{N_R  \times N_R }^L } \\
   {{\bf{Z}}_{N_S  \times N_T } } & {{\bf{Z}}_{N_S  \times N_R } } \\
 \end{array} } \right. \nonumber \\
 &\left. {\begin{array}{*{20}c}
   {{\bf{Z}}_{N_T  \times N_S } }  \\
   {{\bf{Z}}_{N_R  \times N_S } }  \\
   {{\bf{Z}}_{N_S  \times N_S }  + {\bf{Z}}_{N_S  \times N_S }^L }  \\
 \end{array} } \right]
 \left[ {\begin{array}{*{20}c}
   {{\bf{I}}_{N_T } }  \\
   {{\bf{I}}_{N_R } }  \\
   {{\bf{I}}_{N_S } }  \\
 \end{array} } \right] \label{eq:Matrix-2}
\end{align}
Here, $N_T$ is the number of segments of transmitting antennas, $N_R$ is the number of segments of receiving antennas, and $N_S$ is the number of segments of scatterers with variable load impedance (i.e. $N_T + N_R + N_S =N$). 
${\mathbf{Z}}_{i  \times j}$ is $i \times j$ block impedance matrix, where $i=N_T, N_R, N_S$ and $j=N_T, N_R, N_S$. 
${\mathbf{Z}}_{N_T  \times N_T }^L$ is a $N_T  \times N_T$ diagonal matrix corresponding to transmitting antennas and includes their load impedance connected to the ports. 
${\mathbf{Z}}_{N_R  \times N_R }^L$ is a $N_R  \times N_R$ diagonal matrix corresponding to receiving antennas and includes their load impedance connected to the ports. 
${\mathbf{Z}}_{N_S  \times N_S }^L$ is a $N_S  \times N_S$ diagonal matrix corresponding to scatterers with variable load impedance and includes their load impedance connected to the ports. 
${\mathbf{I}}_{N_T }$, ${\mathbf{I}}_{N_R }$, and ${\mathbf{I}}_{N_S }$ are $N_T$, $N_{R}$ and $N_S$ dimensional block current vectors. 
${\mathbf{V}}_{N_T }$ is a $N_T$ dimensional block voltage vector corresponding to the segments of the transmitting antennas. 
A voltage-voltage MIMO channel matrix, compliant to that used in wireless system optimization \cite{QMdR1,ADD1}, is derived from the transfer matrix partitioning 
introduced in (\ref{eq:Matrix-2}), which is based on \cite{MH1}, and described in Appendix A. 
It configures as 
\begin{align}
{\bf{V}}_{N_R } &= {\bf{Z}}_{N_R  \times N_R }^L {\bf{I}}_{N_R } \label{eq:Matrix-A14} \\
                  &= {\bf{H}}^{f}_{N_R  \times N_T } {\bf{V}}_{N_T } \label{eq:Matrix-A15}
\end{align}
with
\begin{align}
{\bf{H}}^{f}_{N_R  \times N_T } &= {\bf{Z}}_{N_R  \times N_R }^L ({\bf{U}}_{N_R  \times N_R }  - {\bf{Y}}_{N_R  \times N_S } {\bf{Z}}_{N_S  \times N_S }^L \nonumber \\ 
                                   &\quad ({\bf{U}}_{N_S  \times N_S } + {\bf{Y}}_{N_S  \times N_S } {\bf{Z}}_{N_S  \times N_S }^L )^{ - 1} {\bf{Y}}_{N_S  \times N_R } \nonumber \\
                                   &\quad {\bf{Z}}_{N_R  \times N_R }^L )^{ - 1}({\bf{U}}_{N_R  \times N_R } + {\bf{Y}}_{N_R  \times N_R } {\bf{Z}}_{N_R  \times N_R }^L )^{ - 1} \nonumber \\
                                   &\quad ({\bf{Y}}_{N_R  \times N_T }  - {\bf{Y}}_{N_R  \times N_S } {\bf{Z}}_{N_S  \times N_S }^L ({\bf{U}}_{N_S  \times N_S }  \nonumber \\ 
                                   &\quad + {\bf{Y}}_{N_S  \times N_S } {\bf{Z}}_{N_S  \times N_S }^L )^{ - 1} {\bf{Y}}_{N_S  \times N_T } ) \label{eq:Matrix-A16}
\end{align} 
It is worth noticing that ${\bf{V}}_{N_R }$ is a receiving voltage vector and entries of ${\bf{H}}^{f}_{N_R  \times N_T }$ - corresponding to the ports of the transmitting/receiving antennas -
are extracted to obtain complex-valued channel gains \cite{NJS1}.
The channel matrix obtained from (\ref{eq:Matrix-A16}) is a full-wave expression as no approximation is introduced through its derivation. 
Finally, we have verified that the channel matrix obtained from (\ref{eq:Matrix-A16}) perfectly agrees with that obtained by full-wave analysis. 
Interestingly, the impedance formulation in (\ref{eq:Matrix-A16}) does not configure as a cascaded model and it presents inverse operators that cannot be 
expanded in Neumann series since their spectral radius is not smaller than the unity. 
Hence, we support the observation done in \cite{AMSASPdH1} concerning the mathematical structure of the scattering formulation, whose 
spectral radius is smaller than the unity thus allowing for a Born series type expansion, 
while the impedance formulation does not allow for such an expansion. 

\subsection{Reduced model}
Here, it is assumed that effect of mutual coupling among the transmitting antennas, receiving antennas, and the RIS is negligible when their current distribution is obtained (The assumption is reasonable when they are in far-field region.). 
According to the assumption, current distribution of the transmitting antennas are approximately obtained from Eq. (\ref{eq:Matrix-2}) as follows. 
\begin{align}
{\bf{V}}_{N_T }  &= ({\bf{Z}}_{N_T  \times N_T }  + {\bf{Z}}_{N_T  \times N_T }^L ){\bf{I}}_{N_T }  + {\bf{Z}}_{N_T  \times N_R } {\bf{I}}_{N_R }  \nonumber \\
                   &\quad + {\bf{Z}}_{N_T  \times N_S } {\bf{I}}_{N_R } \nonumber \\
                   &\approx ({\bf{Z}}_{N_T  \times N_T }  + {\bf{Z}}_{N_T  \times N_T }^L ){\bf{I}}_{N_T }  \nonumber \\
  {\bf{I}}_{N_T } &\approx ({\bf{Z}}_{N_T  \times N_T }  + {\bf{Z}}_{N_T  \times N_T }^L )^{ - 1} {\bf{V}}_{N_T } \label{eq:Matrix-3}
\end{align}
Next, current distribution of the RIS is obtained.
According to the third column of Eq. (\ref{eq:Matrix-2}), current distribution of RIS is approximately obtained as follows. 
\begin{align}
  0                 &= {\bf{Z}}_{N_S  \times N_T } {\bf{I}}_{N_T }  + {\bf{Z}}_{N_S  \times N_R } {\bf{I}}_{N_R }  + ({\bf{Z}}_{N_S  \times N_S }  + {\bf{Z}}_{N_S  \times N_S }^L ){\bf{I}}_{N_S } \nonumber \\
                    &\approx {\bf{Z}}_{N_S  \times N_T } {\bf{I}}_{N_T }  + ({\bf{Z}}_{N_S  \times N_S }  + {\bf{Z}}_{N_S  \times N_S }^L ){\bf{I}}_{N_S } \nonumber \\
  {\bf{I}}_{N_S }  &=  - ({\bf{Z}}_{N_S  \times N_S }  + {\bf{Z}}_{N_S  \times N_S }^L )^{ - 1} {\bf{Z}}_{N_s  \times N_T } {\bf{I}}_{N_T } \nonumber \\
                    &=  - ({\bf{Z}}_{N_S  \times N_S }  + {\bf{Z}}_{N_S  \times N_S }^L )^{ - 1} {\bf{Z}}_{N_s  \times N_T } \nonumber \\ 
                    &\quad ({\bf{Z}}_{N_T  \times N_T }  + {\bf{Z}}_{N_T  \times N_T }^L )^{ - 1} {\bf{V}}_{N_T } \quad (\because {\rm Eq}. (\ref{eq:Matrix-3})) \label{eq:Matrix-4}
\end{align}
After that, current distribution of the receiving antennas is approximately obtained from the second column of Eq. (\ref{eq:Matrix-2}) with the help of Eqs. (\ref{eq:Matrix-3}) and (\ref{eq:Matrix-4}). 
\begin{align}
  0 &= {\bf{Z}}_{N_R  \times N_T } {\bf{I}}_{N_T }  + ({\bf{Z}}_{N_R  \times N_R }  + {\bf{Z}}_{N_R  \times N_R }^L ){\bf{I}}_{N_R } \nonumber \\ 
  &\quad + {\bf{Z}}_{N_R  \times N_S } {\bf{I}}_{N_S } \nonumber \\
  {\bf{I}}_{N_R }  &=  - ({\bf{Z}}_{N_R  \times N_R }  + {\bf{Z}}_{N_R  \times N_R }^L )^{ - 1} \nonumber \\ 
   &\quad ({\bf{Z}}_{N_R  \times N_T } {\bf{I}}_{N_T }  + {\bf{Z}}_{N_R  \times N_S } {\bf{I}}_{N_S } ) \nonumber \\
   &=  - ({\bf{Z}}_{N_R  \times N_R }  + {\bf{Z}}_{N_R  \times N_R }^L )^{ - 1} ({\bf{Z}}_{N_R  \times N_T } ({\bf{Z}}_{N_T  \times N_T } \nonumber \\ 
   &\quad + {\bf{Z}}_{N_T  \times N_T }^L )^{ - 1} {\bf{V}}_{N_T } - {\bf{Z}}_{N_R  \times N_S } ({\bf{Z}}_{N_S  \times N_S }  + {\bf{Z}}_{N_S  \times N_S }^L )^{ - 1} \nonumber \\ 
   &{\bf{Z}}_{N_s  \times N_T } ({\bf{Z}}_{N_T  \times N_T }  + {\bf{Z}}_{N_T  \times N_T }^L )^{ - 1} {\bf{V}}_{N_T } ) \nonumber \\
   &= ({\bf{Z}}_{N_R  \times N_R }  + {\bf{Z}}_{N_R  \times N_R }^L )^{ - 1} ( - {\bf{Z}}_{N_R  \times N_T }  + {\bf{Z}}_{N_R  \times N_S } \nonumber \\
   &\quad ({\bf{Z}}_{N_S  \times N_S }  + {\bf{Z}}_{N_S  \times N_S }^L )^{ - 1} {\bf{Z}}_{N_s  \times N_T } ) \nonumber \\
   &\quad \times ({\bf{Z}}_{N_T  \times N_T }  + {\bf{Z}}_{N_T  \times N_T }^L )^{ - 1} {\bf{V}}_{N_T } \label{eq:Matrix-5}
\end{align}
The receiving voltage vector ${\bf V}_{N_R}$ of the receiving antennas is expressed as follows. 
\begin{equation}
  {\bf{V}}_{N_R }  = {\bf{Z}}_{N_R  \times N_R }^L {\bf{I}}_{N_R } 
                     = {\bf{H}}_{N_R  \times N_T } {\bf{V}}_{N_T } \label{eq:Matrix-7}
\end{equation}
where the voltage-voltage MIMO channel matrix is obtained as
\begin{align}
  {\bf{H}}_{N_R  \times N_T } & = {\bf{Z}}_{N_R  \times N_R }^L ({\bf{Z}}_{N_R  \times N_R }  + {\bf{Z}}_{N_R  \times N_R }^L )^{ - 1}  \nonumber \\ 
  						& \quad \times ( - {\bf{Z}}_{N_R  \times N_T } + {\bf{Z}}_{N_R  \times N_S } \,\, \Phi_{S} \,\, {\bf{Z}}_{N_s  \times N_T } ) \nonumber \\ 
                    				& \quad \times ({\bf{Z}}_{N_T  \times N_T }  + {\bf{Z}}_{N_T  \times N_T }^L )^{ - 1} {\bf{V}}_{N_T } \label{eq:Matrix-7a}
\end{align}
and the RIS state matrix is given by the admittance matrix 
\begin{equation}\label{eqn:RIS_state} 
	\Phi_{S} = ({\bf{Z}}_{N_S  \times N_S}  + {\bf{Z}}_{N_S  \times N_S }^L )^{ - 1}
\end{equation}
which is consistent with the communication model in \cite{GD1} in the far-field approximation, that has been used in sum-rate optimization \cite{ADD1}.
Because the block load impedance matrix ${\bf{Z}}_{N_R  \times N_R }^L$ is a diagonal matrix, its entries are zero except for the ports of the receiving antennas. 
Finally, entries of ${\bf{H}}_{N_R  \times N_T }$ corresponding to the ports of the transmitting/receiving antennas are extracted and the channel matrix is obtained. 

\subsection{Structural scattering}
In NLOS between transmit and receive array, the cascaded communication model in \cite{GD1} vanishes if all the dipole ports are in open circuit, 
on account of the minimum scattering assumption \cite{KK1}.
Importantly, the MoM based formulation leads to a non-zero RIS state matrix when unit cells are unloaded. 
The reason lies in the presence of passive segments that capture the structural scattering of the bare meta-surface structure.
This can be seen by considering the active impedance of a single unit cell of the RIS. 
Suppose we have a linear antenna discretised by three adjacent segments, of which only one embeds port terminals. 
Adopting the MoM formalism, the RIS state matrix for the three segments is found from 
\begin{equation}\label{eqn:ZSStest}
{\bf{Z}}_{N_S  \times N_S}  = 
\begin{pmatrix}
Z & Z_c & Z_c \\
Z_c & Z_p & Z_c \\
Z_c & Z_c & Z 
\end{pmatrix}
\end{equation}
where, without loss of generality, mutual coupling impedance $Z_c$ has been assumed uniform within the macro-pixel, 
with input impedance at the port segment $Z_p$ and at the passive segments $Z$.
The load impedance at the port segment reads
\begin{equation}\label{eqn:ZRIStest}
{\bf{Z}}_{N_S  \times N_S }^L = 
\begin{pmatrix}
0 & 0 & 0 \\
0 & Z_L & 0 \\
0 & 0 & 0 
\end{pmatrix}
\end{equation}
Inserting (\ref{eqn:ZSStest}) and (\ref{eqn:ZRIStest}) in (\ref{eqn:RIS_state}) yields the matrix inversion
\begin{equation}\label{eqn:PHItest}
\Phi_{S} = 
{\begin{pmatrix}
Z & Z_c & Z_c \\
Z_c & Z_p+Z_L & Z_c \\
Z_c & Z_c & Z
\end{pmatrix}}^{-1}
\end{equation}
which gives, in the open circuit ($oc$) case, i.e., $Z_L \rightarrow \infty$
\begin{equation}\label{eqn:PHItest2}
\Phi_{S}^{oc} = 
\begin{pmatrix}
\frac{Z}{Z^2 - Z_c^2} & 0 & \frac{Z_c}{Z^2 - Z_c^2} \\
0 & 0 & 0 \\
\frac{-Z_c}{Z^2 - Z_c^2} & 0 & \frac{Z}{Z^2 - Z_c^2}
\end{pmatrix}
\end{equation}
that encodes the passive scattering of the planar  structure.
Note that differently from \cite{GD1}, the RIS state admittance does not go to zero for unloaded unit cells, 
which was a consequence of the CMS approximation.
\section{Numerical results}
\subsection{Design of RIS}
\begin{figure}[t]
\begin{center}
\includegraphics[width=8.5cm]{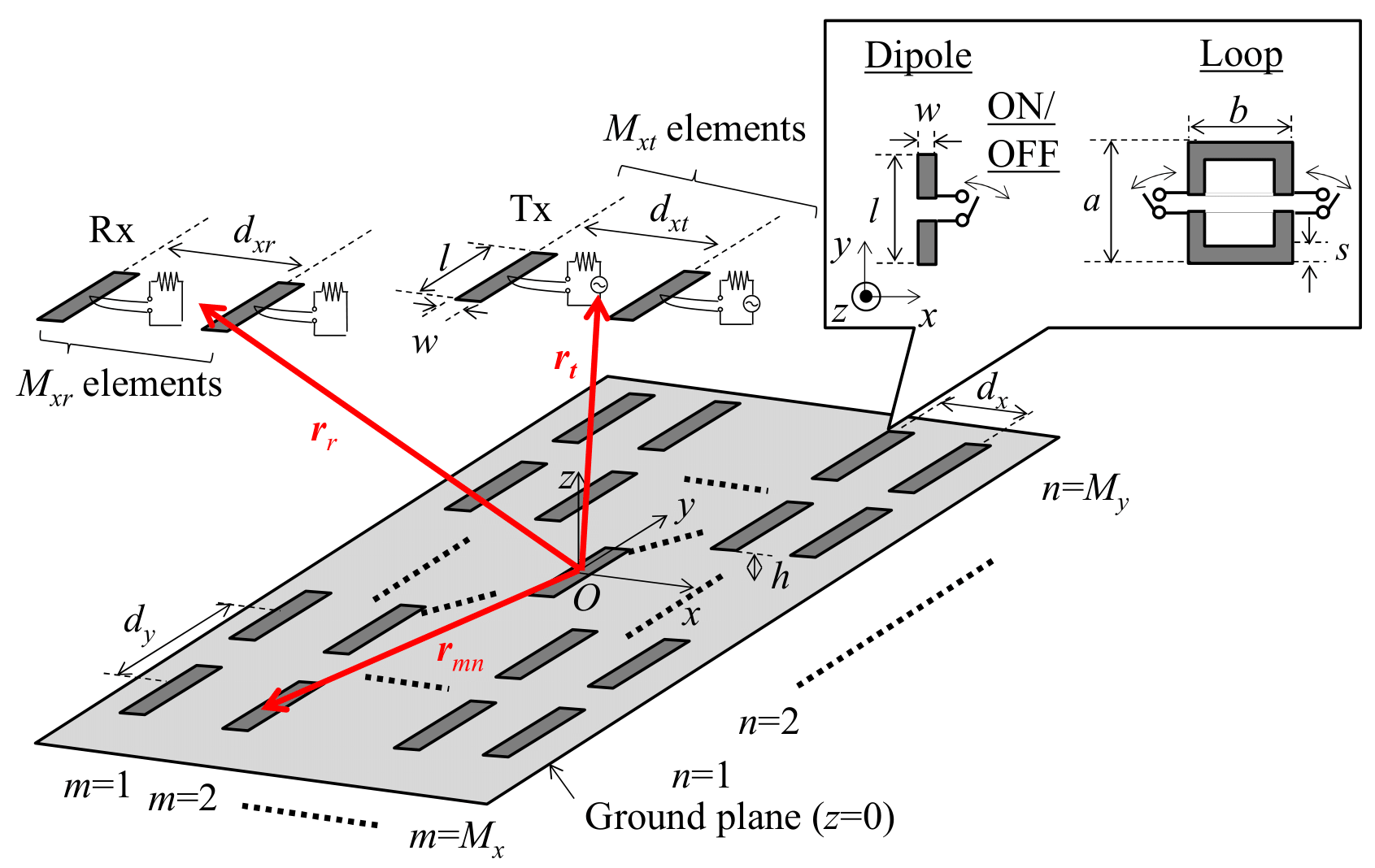}\\
\caption{Geometry of RIS}
\label{fig:RIS}
\end{center}
\end{figure}
\begin{figure}[t]
\begin{center}
\includegraphics[width=8.5cm]{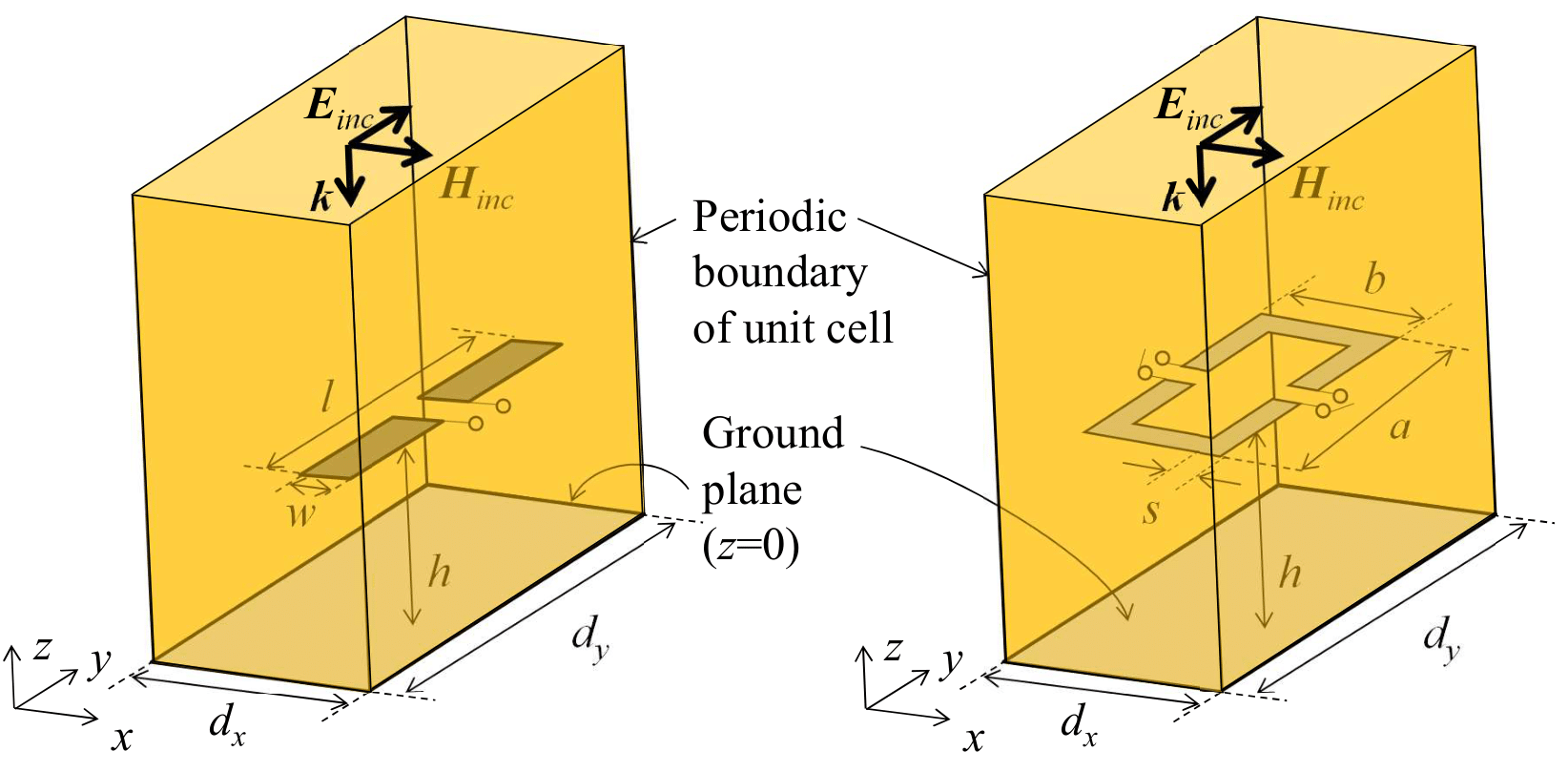}\\
\caption{Unitcell model of infinite array of RIS elements over PEC ground plane (Left: Planar dipole elements, Right: Rectangular loop elements). }
\label{fig:Unitcell}
\end{center}
\end{figure}
Here, it is demonstrated that our proposed expression works for calculating channel capacity between transmitting and receiving antennas in the presence of RIS. 
Geometry of the RIS is shown in Fig. \ref{fig:RIS}. 
Here, either of planar dipole element or planar loop element is used as RIS element. 
Phase of scattering field of the RIS elements is tunable by variable load impedances. 
Variable load impedances loaded with RIS elements are simply modeled as open or short terminations, i.e. the RIS is a so-called single bit RIS.  
Transmitting and receiving antennas are planar dipole array antennas. 
Origin of the coordinate system is correspnding to the center of the RIS and a position vector pointing the center of the $mn$th RIS element is ${\bf r}_{mn}$. 
A position vector pointing the center of the transmitting antenna is ${\bf r}_{t}$ and that pointing the center of the receiving antenna is ${\bf r}_{r}$. 
All of the antenna elements and RIS elements are discretized into triangular segments and their current distribution is expanded using Rao-Wilton-Glisson (RWG) basis function \cite{RWG1}. 

In advance of designing the RIS, phase of reflection coefficient of the RIS elements is obtained numerically. 
Here, method of moments based on periodic Green's function (PGF) is used for numerical analysis of the RIS elements in an infinite periodic array \cite{PMM-1}, \cite{PMM-2}.  
Poor convergence of the PGF is enhanced by introducing Ewald transformation with the optimum splitting parameter \cite{PMM-3}-\cite{PMM-5}. 
Singularity at a point where the source/observation points are is overlapping is annihilated using L’Hospital rule \cite{PMM-6}, \cite{PMM-7}. 

\begin{figure}[t]
\begin{center}
\includegraphics[width=8.5cm]{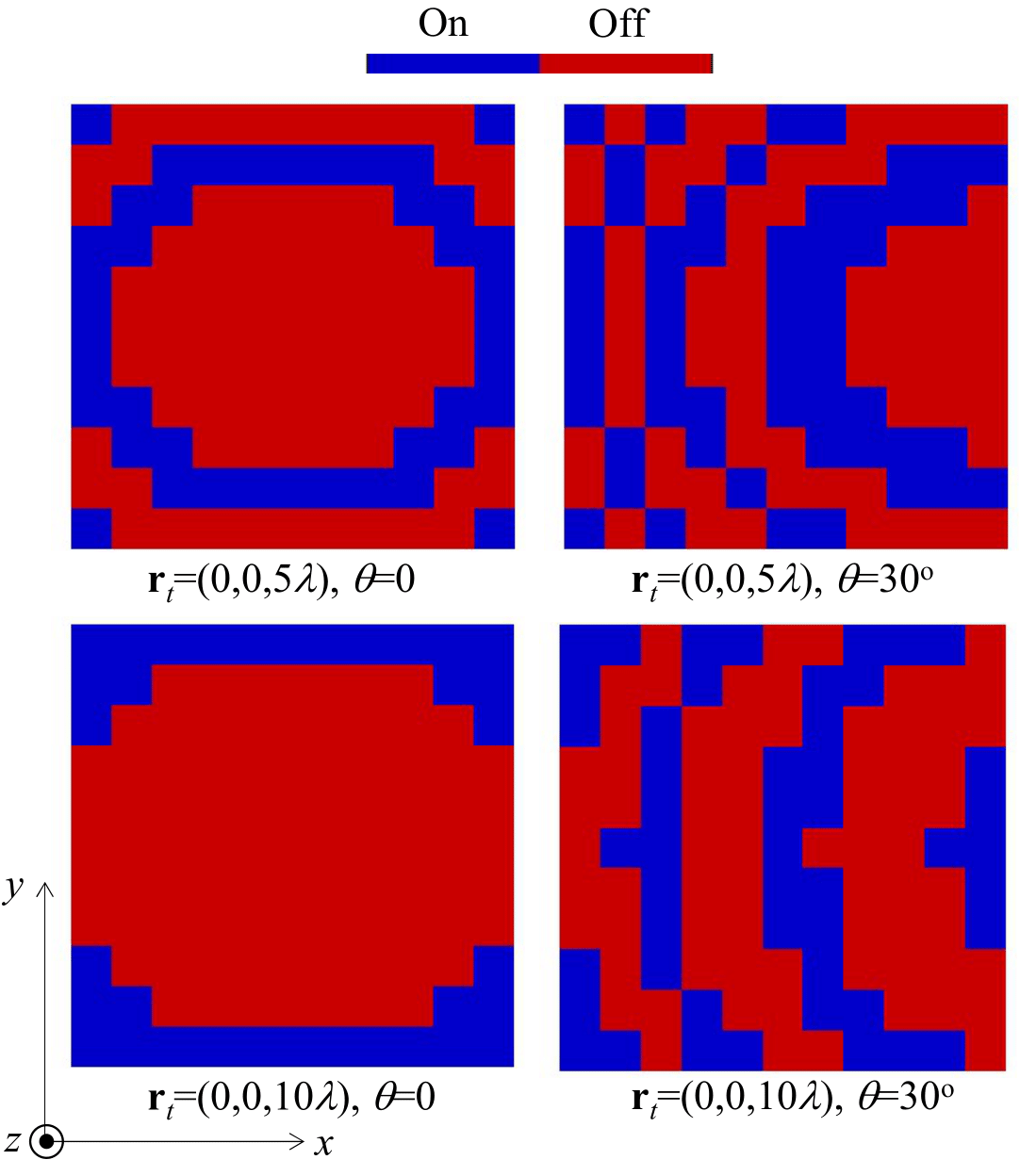}\\
\caption{On/off states of RIS elements composed of dipole elements. }
\label{fig:On-off-states_dipole}
\end{center}
\end{figure}
\begin{figure}[t]
\begin{center}
\includegraphics[width=8.5cm]{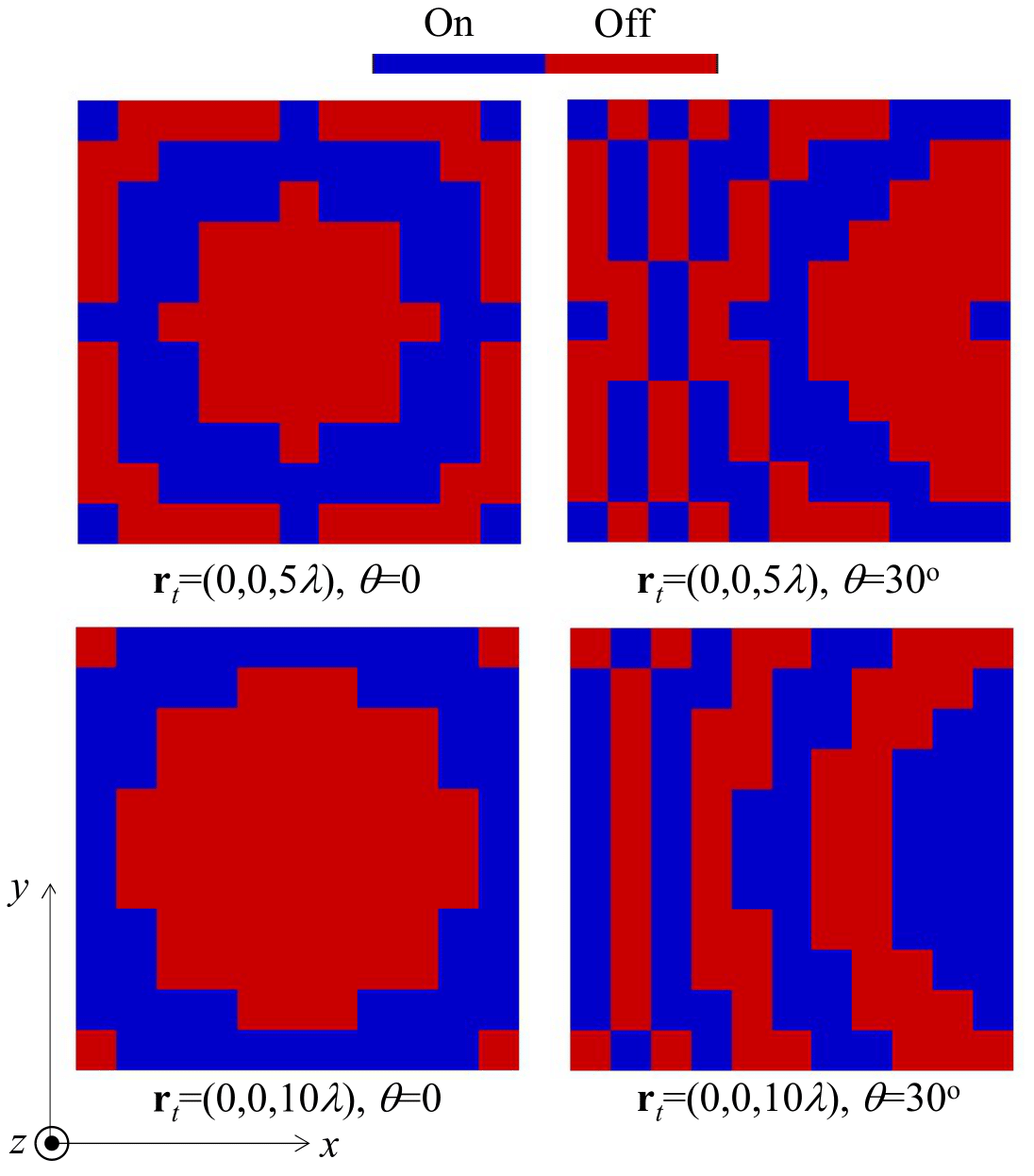}\\
\caption{On/off states of RIS elements composed of loop elements. }
\label{fig:On-off-states_loop}
\end{center}
\end{figure}
\begin{figure}[t]
\begin{center}
\includegraphics[width=8.5cm]{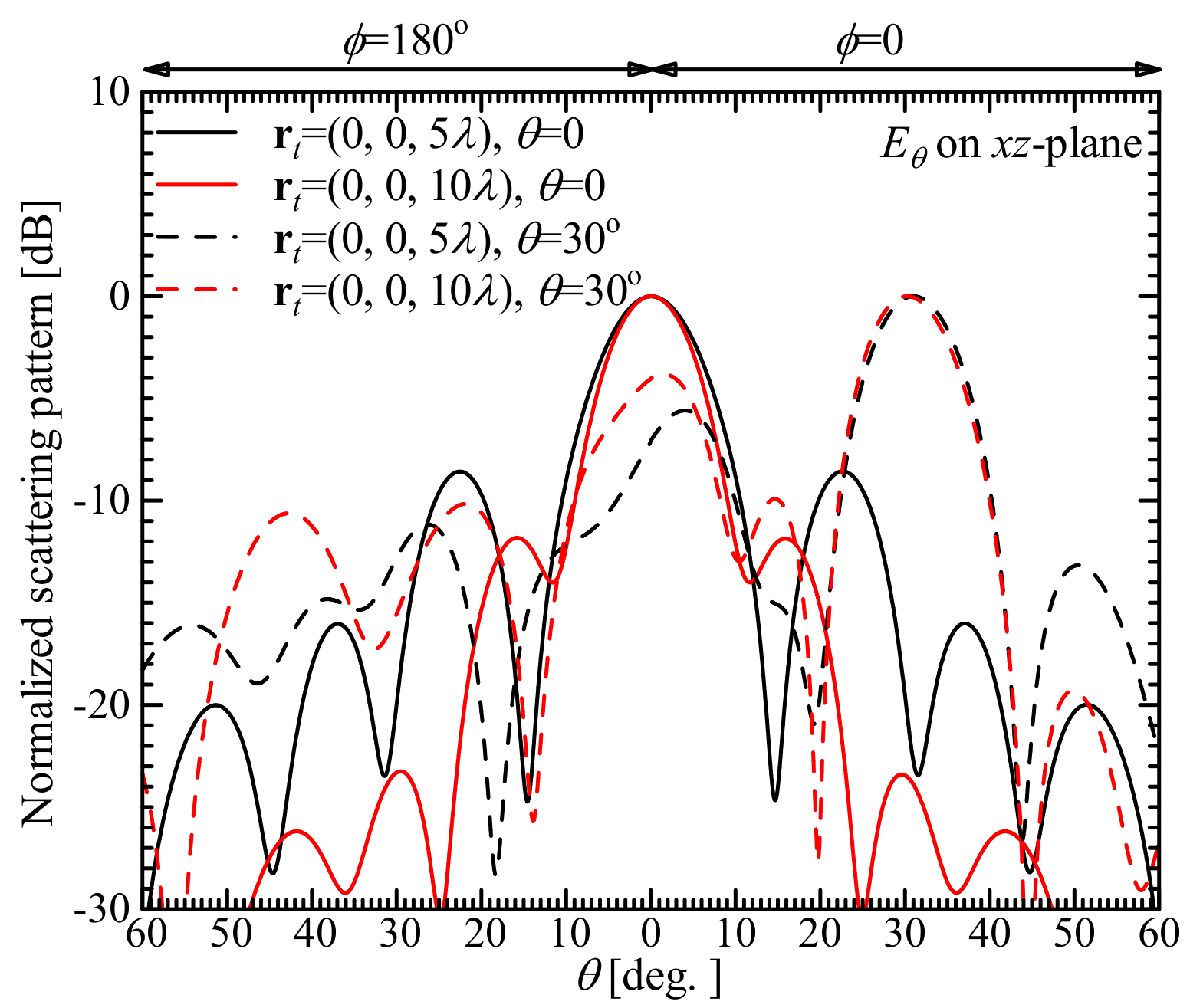}\\
\caption{Scattering patterns of RIS composed of dipole elements. }
\label{fig:Scattering patterns_dipole}
\end{center}
\end{figure}
\begin{figure}[t]
\begin{center}
\includegraphics[width=8.5cm]{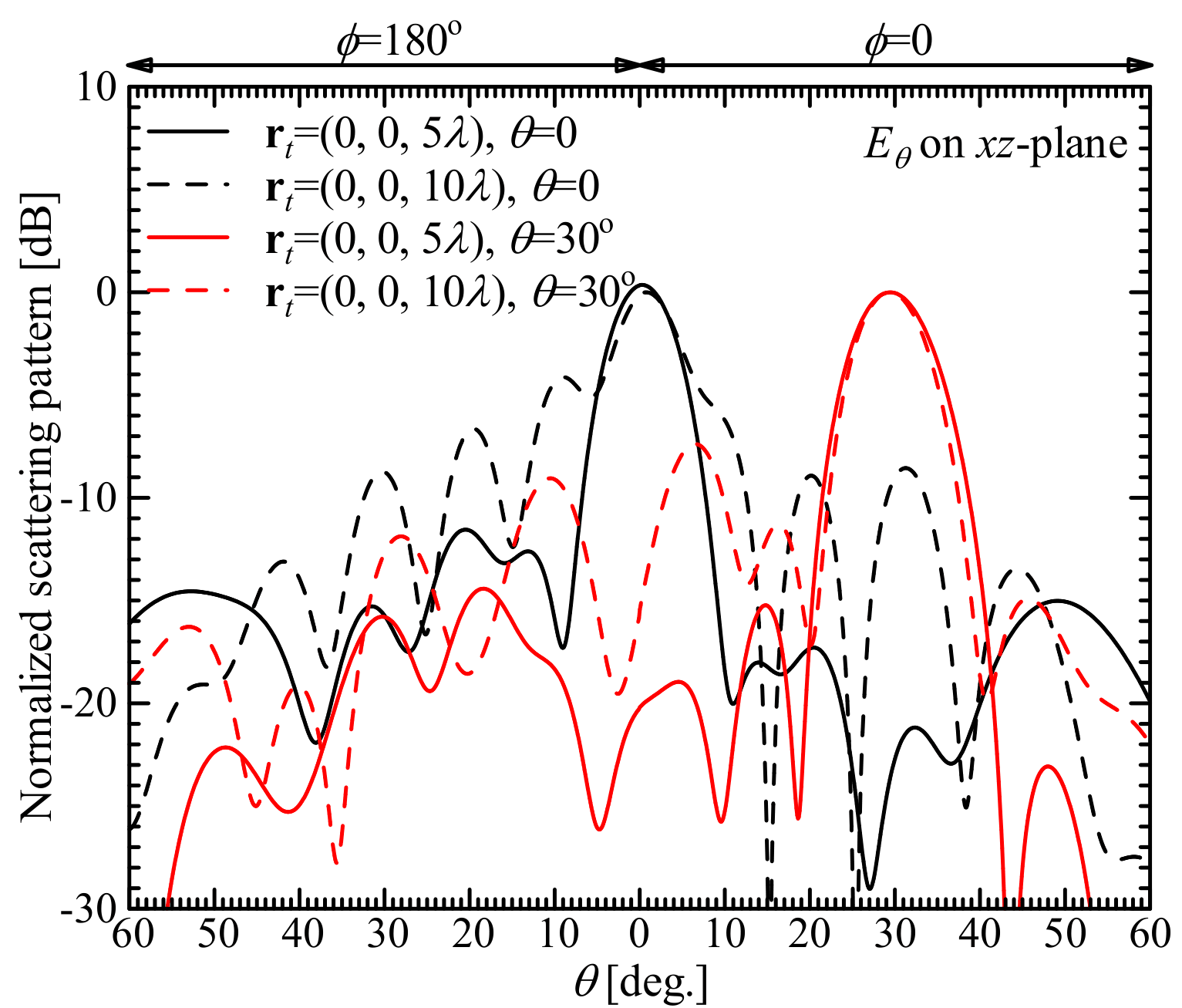}\\
\caption{Scattering patterns of RIS composed of loop elements. }
\label{fig:Scattering patterns_loop}
\end{center}
\end{figure}

The geometry of unit cell model composed of planar dipole elements and planar loop elements is shown in Fig. \ref{fig:Unitcell}. 
Phase of reflection coefficient of the unit cell model is obtained for plane wave of normal incidence and is shown in Table \ref{table:On-off-states}. 
According to the Table \ref{table:On-off-states}, it is found that phase shift of 180 deg. is approximately available by switching variable load impedances from on to off, and vice versa. 
Therefore, it can be said that the planar dipole elements and planar loop elements with variable load impedances are applicable to a single-bit RIS. 

The single-bit RIS is designed using the planar dipole elements and planar loop elements. 
Based on the phase of reflection coefficient, on/off states of the RIS elements are designed so that the scattering fields of the RIS elements are superposed with the minimum error at specific main beam direction. 
Here, dimensions of the RIS and transmitting antennas are shown in TABLE \ref{table:Dimensions_RIS}. 
The RIS is designed so that its main beam is directed to $\theta=0, 30^\circ$. 
The RIS is analysed using the MoM with the layered media Green's function(LMGF) and infinite ground plane of the RIS is modeled using the LMGF \cite{LMGF1}. 
A direct wave component is extracted from the LMGF and calculated in the spatial domain so that the convergence of the LMGF is enhanced \cite{LMGF2}, \cite{LMGF3}. 
Singularity at a point where the source/observation points are is overlapping is annihilated using coordinate transformation and analytic integral \cite{RWG2}, \cite{RWG3}. 
So-called Sommerfeld integral in the LMGF is performed efficiently via numerical interpolation by using Taylor expansions \cite{LMGF4}, \cite{LMGF5}. 

On/off states of the RIS elements are shown in Figs. \ref{fig:On-off-states_dipole} and \ref{fig:On-off-states_loop}. 
It is found that on/off states of the RIS elements are switched corresponding to the main beam direction or position of the transmitting antenna. 
The typical phase mask of a lens emerges upon RIS unit cell optimization in both dipole and loop based reflecting sufaces. 
Normalised scattering patterns of the RISs composed of the planar dipole elements and the planar loop elements are shown in Figs. \ref{fig:Scattering patterns_dipole} and \ref{fig:Scattering patterns_loop}, respectively. 
It is found that main beam of the RIS is directed to the specific angle because on/off states of the RIS elements are switched so that their scattering fields are with the minimum error at the direction. 	

\begin{table*}[t]
\caption{Phase of reflection coefficient of planar dipole and planar loop elements.}
\label{table:On-off-states}
\centering
\begin{tabular}{l|c|c|c}
\hline \hline
\multicolumn{2}{c|}{}								& Planar dipole 						& Planar loop							\\
\multicolumn{2}{c|}{Type of elements}					& ($l=0.5\lambda, w=0.01\lambda, $		& ($a=b=0.4\lambda, w=0.01\lambda, $	 	\\
\multicolumn{2}{c|}{and their dimensions}				& $h=0.25\lambda, d_x=0.5\lambda, $		& $h=0.25\lambda, d_x=d_y=0.6\lambda)$ 	\\
\multicolumn{2}{c|}{}								& $d_y=0.7\lambda)$					&									\\ \hline
Phase of 				&	On state 					&	 6.4								& -56								\\
reflection				&	(Short termination)			&									&									\\ \cline{2-4}
coefficient [deg.]		&	Off state					& 164.6								& 122.8            				 		\\ 
					&	(Open termination)			&									&									\\ \hline
\end{tabular}
\end{table*}
\begin{table*}[t]
\caption{Dimensions of the RIS and transmitting/receiving antennas. }
\label{table:Dimensions_RIS}
\centering
\begin{tabular}{c|c|c|c|c|c|c|c}
\hline \hline
\multicolumn{4}{c|}{RIS}															& \multicolumn{2}{c|}{Tx antenna}			& \multicolumn{2}{c}{Rx antenna}    	\\ \cline{1-4}
\multicolumn{2}{c|}{Planar dipole elements}		& \multicolumn{2}{c|}{Planar loop elements}	& \multicolumn{2}{c|}{}					& \multicolumn{2}{c}{}				\\ \hline
$l$			&	$0.5\lambda$				& $a=b$		& $0.4\lambda$			& $l$			& $0.5\lambda$		& $l$			& $0.5\lambda$					\\ \hline
$w$			&	$0.01\lambda$				& $w$		& $0.01\lambda$			& $w$			& $0.01\lambda$		& $w$			& $0.01\lambda$					\\ \hline
$d_x$		&	$0.5\lambda$				& $d_x$		& $0.6\lambda$			& $d_{xt}$			& $0.5\lambda$		& $d_{xr}$			& $0.5\lambda$					\\ \hline
$d_y$		&	$0.7\lambda$				& $d_y$		& $0.6\lambda$			& $M_{xt}$		& $2$				& $M_{xr}$		& $2$							\\ \hline
$h$			&	$0.25\lambda$				& $h$		& $0.25\lambda$			& ${\bf r}_{t}$		& $(0,0,5\lambda)$		& ${\bf r}_{r}$		& $(R \cos \theta, 0, R \sin \theta)$ 	\\ \cline{1-4}
$M_x=M_y$	&	$11$					& $M_x=M_y$	& $11$					&				& $(0,0,10\lambda)$ 	&				& $\theta=0, 30^\circ$				\\ \hline
\end{tabular}
\end{table*}
\subsection{Channel capacity}

Here, it is demonstrated that our proposed expression of the channel capacity is applicable to a wireless communication system including RIS. 
Dimensions of the transmitting/receiving antennas and the RIS are shown in Table \ref{table:Dimensions_RIS}. 
Channel capacity is expressed as follows. 
\begin{equation}\label{eqn:C}
C = \log _2 \left| {{\bf{I}} + {\bf{H H}}^\dag  {\frac{\gamma}{M_{xt} }}} \right|
\end{equation}
Here, ${\bf{I}}$ is an unit matrix, ${\bf H}$ is a $M_{tr} \times M_{xr}$ channel matrix extracted from ${\bf{H}}_{N_R  \times N_T }$, and $\gamma$ is signal-to-noise ratio (i.e. Ratio between transmitting power and noise power at receiving antennas. ) 

\begin{figure}[!htb]
\begin{center}
\includegraphics[width=8.5cm]{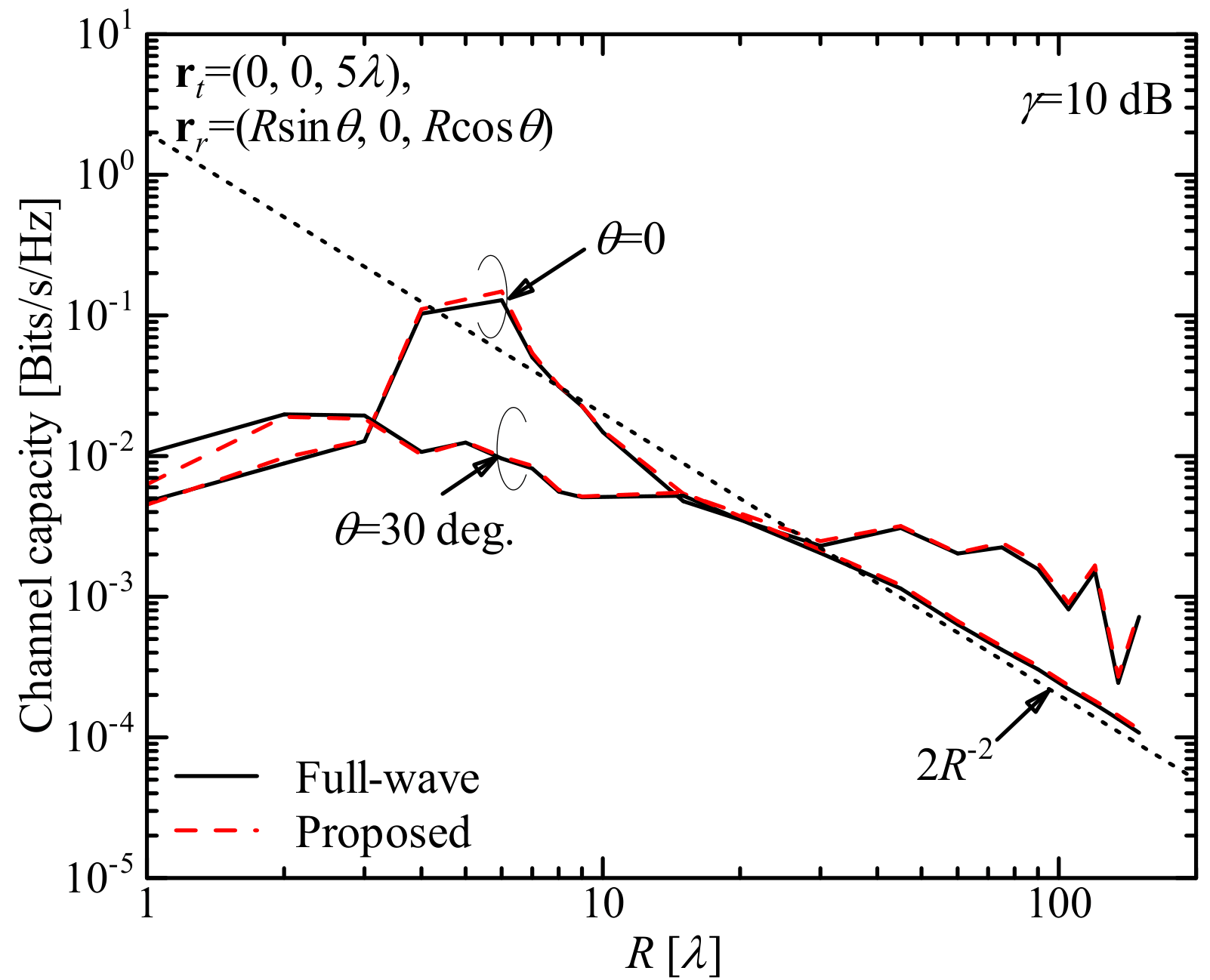}\\
\caption{Channel capacity of RIS composed of planar dipole elements (${\bf r}_{t}=(0, 0, 5\lambda)$). }
\label{fig:RIS_dipole_5lambda}	
\end{center}
\end{figure}

\begin{figure}[!htb]
\begin{center}
\includegraphics[width=8.5cm]{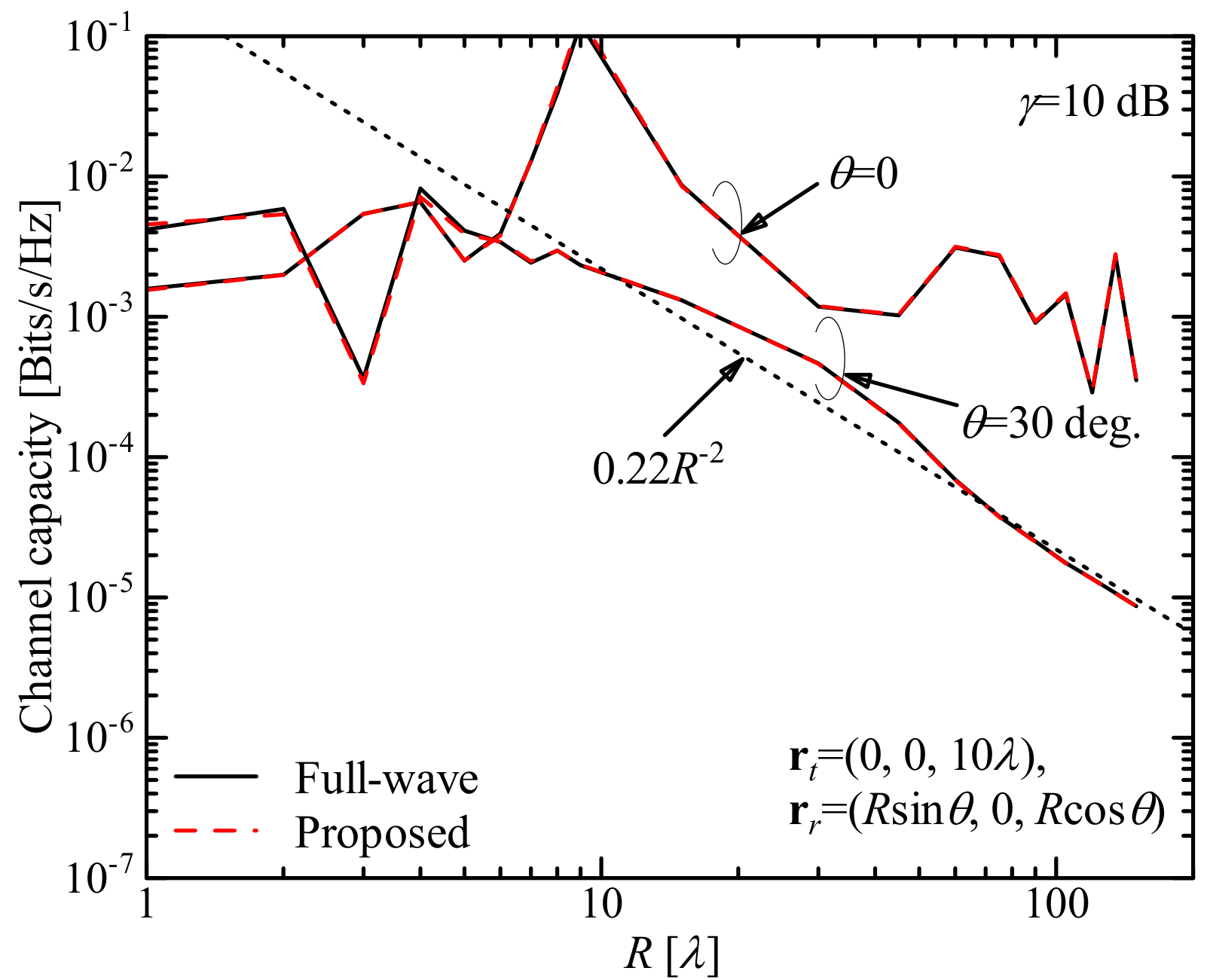}\\
\caption{Channel capacity of RIS composed of planar dipole elements  (${\bf r}_{t}=(0, 0, 10\lambda)$). }
\label{fig:RIS_dipole_10lambda}
\end{center}
\end{figure}

\begin{figure}[!htb]
\begin{center}
\includegraphics[width=8.5cm]{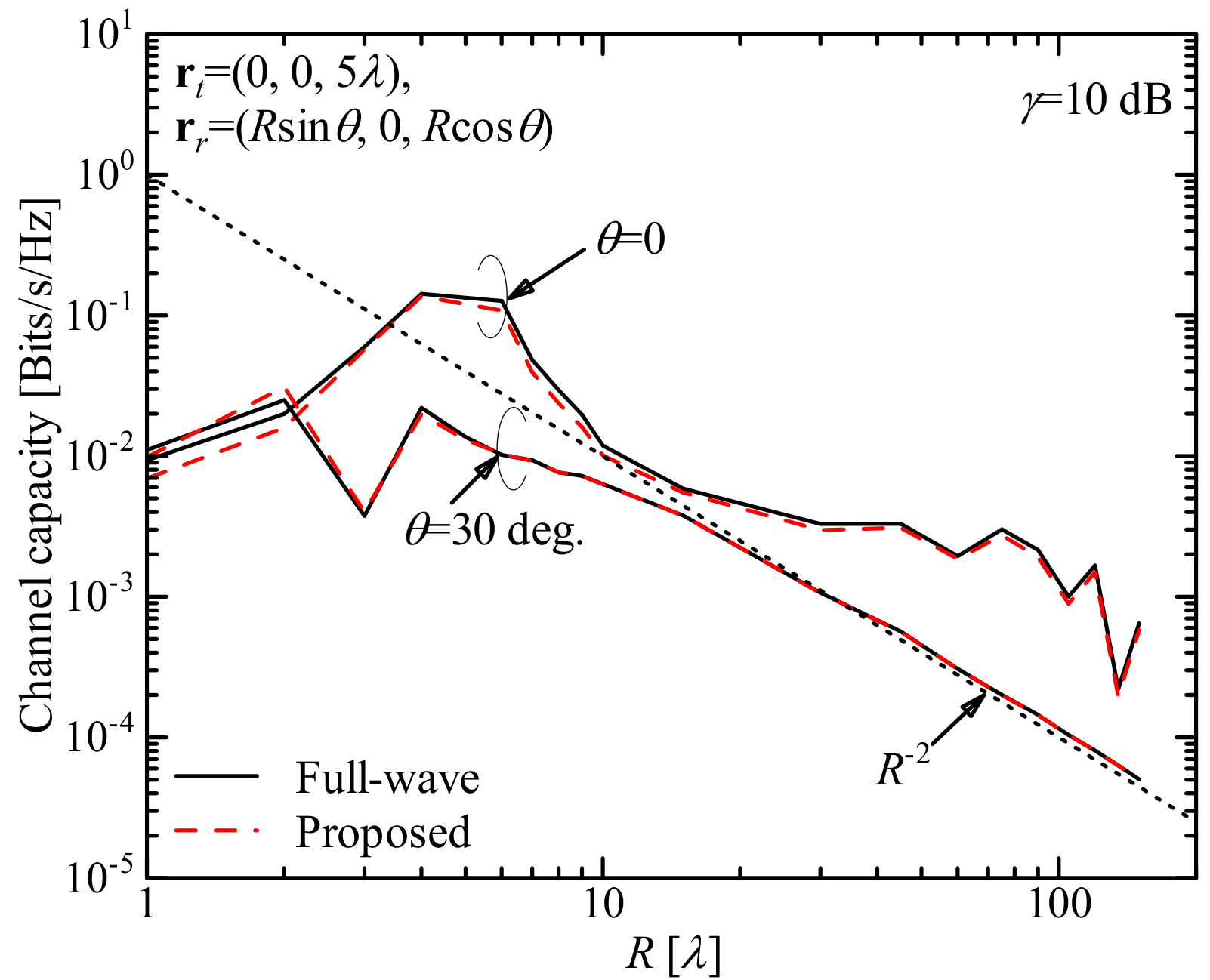}\\
\caption{Channel capacity of RIS composed of planar loop elements (${\bf r}_{t}=(0, 0, 5\lambda)$). }
\label{fig:RIS_loop_5lambda}	
\end{center}
\end{figure}

\begin{figure}[!ht]
\begin{center}
\includegraphics[width=8.5cm]{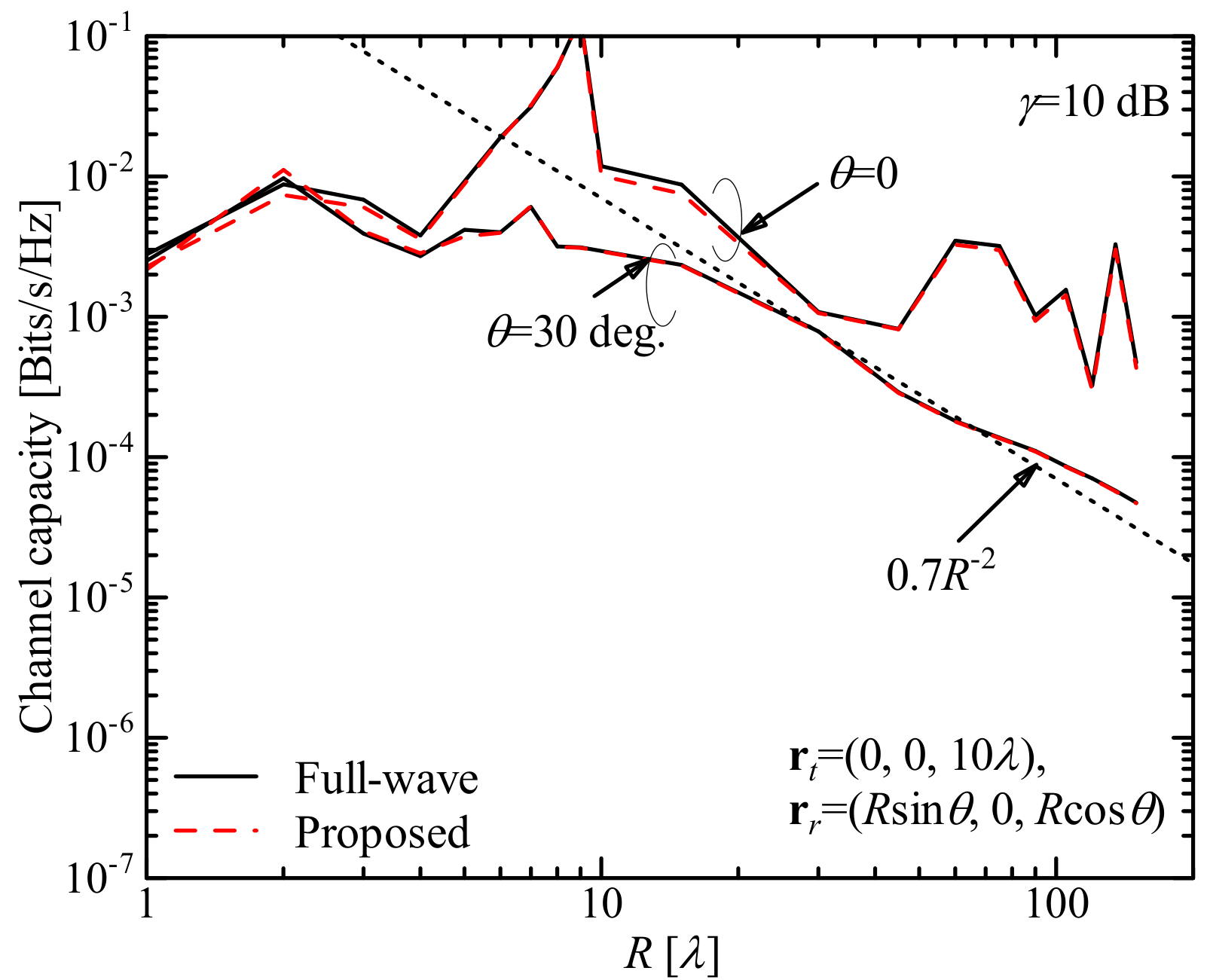}\\
\caption{Channel capacity of RIS composed of planar loop elements  (${\bf r}_{t}=(0, 0, 10\lambda)$). }
\label{fig:RIS_loop_10lambda}
\end{center}
\end{figure}

Simulated channel capacity is shown in Figs. \ref{fig:RIS_dipole_5lambda} and \ref{fig:RIS_dipole_10lambda} for the RIS composed of the planar dipole array antennas. 
Also, simulated channel capacity is shown in Figs. \ref{fig:RIS_loop_5lambda} and \ref{fig:RIS_loop_10lambda} for the RIS composed of the planar loop array antennas. 
It is shown that the channel capacity obtained using the proposed expression agrees well with that obtained using the full-wave analysis except for small error when the receiving antennas are too close to the RIS (e.g. $R=1 \lambda$). 
This is predicted at large surface-receiver distances $R$ and in absence of LOS, for which 
\begin{equation}
  {\bf{Z}}_{N_R  \times N_S } \sim \frac{\textbf{a}_{N_R  \times N_S } }{R},
\end{equation}
as explained in \cite{ASBMH1}, and where $\textbf{a}_{N_R  \times N_S }$ is the array steering vector and upon substitution in (\ref{eq:Matrix-7a}) and (\ref{eqn:C}) yields 
\begin{equation}
 C \sim \left| {{\bf{H H}}^\dag  {\frac{\gamma}{M_{xt} }}} \right| \sim \frac{1}{R^2}.
\end{equation}
Importantly, this is consistent (and provides a link) with path-loss models (see for example \cite{DRDdRT1,GSK1}), and provides an assessment 
of the impedance model for arbitrary RIS structures. 
The small error of the channel capacity obtained using the proposed expression stems from the effect of the mutual coupling among transmitting antennas, 
receiving antennas and the RIS that is neglected in its formulation. 
However, it can be said that applicability of the proposed expression is kept because the receiving antennas are usually far from the RIS in practical system. 

On the other hand, numerical results should be further discussed from the physical point of view. 
The receiving antennas receive direct wave from the transmitting antenna and scattering wave from the RIS simultaneously. 
Therefore, the channel capacity reflects the result of the superposition of the direct/scattering waves. 
For example, as shown in the channel capacity corresponding to the RIS whose angle of main beam is directed to $\theta=0$, gradient of the channel capacity between 
the transmitting/receiving antennas is non-constant when the receiving antennas receive both of the direct/scattering waves. 
The non-constant gradient of the channel capacity comes from phase difference between the direct/scattering waves. 
On the other hand, as shown in the channel capacity corresponding to the RIS whose angle of main beam is directed to $\theta=30^\circ$, 
the gradient of the channel capacity between the transmitting/receiving antennas is proportional to $\frac{1}{R^2}$ when the $R$ becomes large. 
The $\frac{1}{R^2}$ gradient of the channel capacity is found because direct wave from the transmitting antennas is negligible and $R$ approaches to boundary 
of far-field region of the RIS (e.g. $\frac{2D^2}{\lambda} = 148 \lambda$, where $D$ is diagonal line of the RIS composed of the planar dipole elements).
According to the numerical results and discussions, it can be said that the proposed expression is applicable to calculating the channel capacity between 
the transmitting/receiving antennas in the presence of the RIS. 

\section{Conclusion}
\label{sec:Conclusion}
We have obtained a cascaded communication model for RIS-assisted wireless links 
from a first-principle electromagnetic model. 
The formulation has been achieved by partitioning the interaction matrices of the method-of-moments. 
We have found that the model configures as a cascaded product of interaction matrices in the far-field of the RIS unit cell.
This extends communication models based on impedance matrices that has been originally derived 
under the canonical minimum scattering approximation.
The system-level performance of RIS-assisted end-to-end MIMO communications can be accurately simulated with 
Method of Moments based channel matrices. 
Finite-size RIS unit cells with arbitrary geometry and dimensions can be characterised and subsequently optimised 
to achieve prescribed electromagnetic shaping functions.  
Future work is required to devise computationally efficient methods for multilayer and transmissive RIS structures.

\newpage
\appendix{\bf APPENDIX A: Full-wave Expression of Channel Matrix} \label{sec:app}\\
Here, full-wave expression of a channel matrix is derived in the same manner with \cite{Arbitrary-load1}. 
Starting from Eq. (\ref{eq:Matrix-2}), following expression is obtained. 
\begin{align}
  \left[ {\begin{array}{*{20}c}
   {\bf{V}}_{N_T}  \\
   -{\bf{Z}}_{N_R  \times N_R }^L {\bf{I}}_{N_R } \\
   -{\bf{Z}}_{N_S  \times N_S }^L {\bf{I}}_{N_S }  \\
 \end{array} } \right] &= \left[ {\begin{array}{*{20}c}
   {{\bf{Z}}_{N_T  \times N_T }  + {\bf{Z}}_{N_T  \times N_T }^L } \\
   {{\bf{Z}}_{N_R  \times N_T } } & {{\bf{Z}}_{N_R  \times N_R } } \\
   {{\bf{Z}}_{N_S  \times N_T } } & {{\bf{Z}}_{N_S  \times N_R } } \\
 \end{array} } \right. \nonumber \\
 &\quad \left. {\begin{array}{*{20}c}
   {{\bf{Z}}_{N_T  \times N_S } }  \\
   {{\bf{Z}}_{N_R  \times N_S } }  \\
   {{\bf{Z}}_{N_S  \times N_S } }  \\
 \end{array} } \right]
 \left[ {\begin{array}{*{20}c}
   {{\bf{I}}_{N_T } }  \\
   {{\bf{I}}_{N_R } }  \\
   {{\bf{I}}_{N_S } }  \\
 \end{array} } \right] \label{eq:Matrix-AA1} \\
{\bf V}_N^{\prime} &={\bf Z}_{N \times N}^{\prime} {\bf I}_N \label{eq:Matrix-AA2}
\end{align}
Here, ${\bf V}_N^{\prime} \equiv [{\bf{V}}_{N_T}, -{\bf{Z}}_{N_R  \times N_R }^L {\bf{I}}_{N_R }, -{\bf{Z}}_{N_S  \times N_S }^L {\bf{I}}_{N_S }]^{T}$ and  ${\bf Z}_{N \times N}^{\prime}$ is an $N \times N$ impedance matrix where the block load impedance matrix at the transmitting antenna is embedded. 
Eq. (\ref{eq:Matrix-AA2}) can be solved by multiplying ${\mathbf{Y}}_{N \times N} \equiv \{{\bf Z}_{N \times N}^{\prime}\}^{-1}$ with its both sides and following expression is obtained. 
\begin{align}
  {\mathbf{I}}_N  &= {\mathbf{Y}}_{N \times N} {\mathbf{V}}_N^{\prime} \nonumber \\ 
  \left[ {\begin{array}{*{20}c}
   {{\mathbf{I}}_{N_T } }  \\
   {{\mathbf{I}}_{N_R } }  \\
   {{\mathbf{I}}_{N_S } }  \\
 \end{array} } \right] &= \left[ {\begin{array}{*{20}c}
   {{\mathbf{Y}}_{N_T  \times N_T } } & {{\mathbf{Y}}_{N_T  \times N_R } } \\
   {{\mathbf{Y}}_{N_R  \times N_T } } & {{\mathbf{Y}}_{N_R  \times N_R } } \\
   {{\mathbf{Y}}_{N_S  \times N_T } } & {{\mathbf{Y}}_{N_S  \times N_R } } \\
 \end{array} } \right. \nonumber \\
 &\quad \left. {\begin{array}{*{20}c}
   {{\mathbf{Y}}_{N_T  \times N_S } }  \\
   {{\mathbf{Y}}_{N_R  \times N_S } }  \\
   {{\mathbf{Y}}_{N_S  \times N_S } }  \\
 \end{array} } \right]
 \left[ {\begin{array}{*{20}c}
   {{\mathbf{V}}_{N_T } }  \\
   { - {\mathbf{Z}}_{N_R  \times N_R }^L {\mathbf{I}}_{N_R } }  \\
   { - {\mathbf{Z}}_{N_S  \times N_S }^L {\mathbf{I}}_{N_S } }  \\
 \end{array} } \right] \label{eq:Matrix-A2}
\end{align} 
According to Eq. (\ref{eq:Matrix-A2}), following three block matrix equations are obtained. 
\begin{align}
  {}
  {\mathbf{I}}_{N_T }  = {\mathbf{Y}}_{N_T  \times N_T } {\mathbf{V}}_{N_{T} }  - {\mathbf{Y}}_{N_T  \times N_{R} } {\mathbf{Z}}_{N_{R}  \times N_{R} }^L {\mathbf{I}}_{N_{R} } & \nonumber \\ 
                              \quad \quad \quad \quad - {\mathbf{Y}}_{N_T  \times N_{S} } {\mathbf{Z}}_{N_{S}  \times N_{S} }^L {\mathbf{I}}_{N_{S} } \label{eq:Matrix-A3} \\
  {\mathbf{I}}_{N_R }  = {\mathbf{Y}}_{N_R  \times N_T } {\mathbf{V}}_{N_{T} }  - {\mathbf{Y}}_{N_R  \times N_{R} } {\mathbf{Z}}_{N_{R}  \times N_{R} }^L {\mathbf{I}}_{N_{R} } & \nonumber \\  
                              \quad \quad \quad \quad - {\mathbf{Y}}_{N_R  \times N_{S} } {\mathbf{Z}}_{N_{S}  \times N_{S} }^L {\mathbf{I}}_{N_{S} } \label{eq:Matrix-A4} \\
  {\mathbf{I}}_{N_S }  = {\mathbf{Y}}_{N_S  \times N_T } {\mathbf{V}}_{N_{T} }  - {\mathbf{Y}}_{N_S  \times N_{R} } {\mathbf{Z}}_{N_{R}  \times N_{R} }^L {\mathbf{I}}_{N_{R} }  & \nonumber \\ 
                              \quad \quad \quad \quad - {\mathbf{Y}}_{N_S  \times N_{S} } {\mathbf{Z}}_{N_{S}  \times N_{S} }^L {\mathbf{I}}_{N_{S} } \label{eq:Matrix-A5}
\end{align}
(\ref{eq:Matrix-A5}) can be transformed as follows. 
\begin{align}
{\bf{I}}_{N_S }  &= ({\bf{U}}_{N_S  \times N_S }  + {\bf{Y}}_{N_S  \times N_S } {\bf{Z}}_{N_S  \times N_S }^L )^{ - 1} \nonumber \\ 
                   &\quad ({\bf{Y}}_{N_S  \times N_T } {\bf{V}}_{N_T } - {\bf{Y}}_{N_S  \times N_R } {\bf{Z}}_{N_R  \times N_R }^L {\bf{I}}_{N_R }) 
\label{eq:Matrix-A6}
\end{align}
Here, ${\mathbf{U}}_{N_{S}  \times N_{S} }$ is $N_{S} \times N_{S}$ unit matrix. 
The next, Eq. (\ref{eq:Matrix-A4}) can be transformed as follows. 
\begin{align}
{\bf{I}}_{N_R } &= ({\bf{U}}_{N_R  \times N_R }  + {\bf{Y}}_{N_R  \times N_R } {\bf{Z}}_{N_R  \times N_R }^L )^{ - 1} \nonumber \\ 
                  &\quad ( {\bf{Y}}_{N_R  \times N_T } {\bf{V}}_{N_T } - {\bf{Y}}_{N_R  \times N_S } {\bf{Z}}_{N_S  \times N_S }^L {\bf{I}}_{N_S } ) 
\label{eq:Matrix-A7}
\end{align}
Equation (\ref{eq:Matrix-A6}) is substituted into Eq. (\ref{eq:Matrix-A7}) and following expression is obtained. 
\begin{align}
  {\bf{I}}_{N_R }  &= ({\bf{U}}_{N_R  \times N_R }  + {\bf{Y}}_{N_R  \times N_R } {\bf{Z}}_{N_R  \times N_R }^L )^{ - 1} ({\bf{Y}}_{N_R  \times N_T } {\bf{V}}_{N_T } \nonumber \\ 
                    &\quad - {\bf{Y}}_{N_R  \times N_S } {\bf{Z}}_{N_S  \times N_S }^L ({\bf{U}}_{N_S  \times N_S }  + {\bf{Y}}_{N_S  \times N_S } {\bf{Z}}_{N_S  \times N_S }^L )^{ - 1} \nonumber \\ 
                    &\quad ({\bf{Y}}_{N_S  \times N_T } {\bf{V}}_{N_T }  - {\bf{Y}}_{N_S  \times N_R } {\bf{Z}}_{N_R  \times N_R }^L {\bf{I}}_{N_R } )) \label{eq:Matrix-A8} \\
                    &= ({\bf{U}}_{N_R  \times N_R }  + {\bf{Y}}_{N_R  \times N_R } {\bf{Z}}_{N_R  \times N_R }^L )^{ - 1} \left\{ \right. ({\bf{Y}}_{N_R  \times N_T } \nonumber \\
                    &\quad - {\bf{Y}}_{N_R  \times N_S } {\bf{Z}}_{N_S  \times N_S }^L ({\bf{U}}_{N_S  \times N_S }  + {\bf{Y}}_{N_S  \times N_S } {\bf{Z}}_{N_S  \times N_S }^L )^{ - 1} \nonumber \\ 
                    &\quad {\bf{Y}}_{N_S  \times N_T } ) {\bf{V}}_{N_T }  + {\bf{Y}}_{N_R  \times N_S } {\bf{Z}}_{N_S  \times N_S }^L ({\bf{U}}_{N_S  \times N_S } \nonumber \\ 
                    &\quad + {\bf{Y}}_{N_S  \times N_S } {\bf{Z}}_{N_S  \times N_S }^L )^{ - 1} {\bf{Y}}_{N_S  \times N_R } {\bf{Z}}_{N_R  \times N_R }^L {\bf{I}}_{N_R } \left. \right\} \!\!\!\! \label{eq:Matrix-A9}
\end{align}
Equation (\ref{eq:Matrix-A9}) can be solved for ${\bf{I}}_{N_R }$ and ${\bf{I}}_{N_R }$ is obtained as follows. 
\begin{align}
  {\bf{I}}_{N_R }  &= ({\bf{U}}_{N_R  \times N_R }  - {\bf{Y}}_{N_R  \times N_S } {\bf{Z}}_{N_S  \times N_S }^L ({\bf{U}}_{N_S  \times N_S }  + {\bf{Y}}_{N_S  \times N_S } \nonumber \\ 
                    &\quad {\bf{Z}}_{N_S  \times N_S }^L )^{ - 1} {\bf{Y}}_{N_S  \times N_R } {\bf{Z}}_{N_R  \times N_R }^L )^{ - 1} ({\bf{U}}_{N_R  \times N_R } \nonumber \\ 
                    &\quad + {\bf{Y}}_{N_R  \times N_R } {\bf{Z}}_{N_R  \times N_R }^L )^{ - 1} ({\bf{Y}}_{N_R  \times N_T }  - {\bf{Y}}_{N_R  \times N_S } \nonumber \\
                    &\quad {\bf{Z}}_{N_S  \times N_S }^L ({\bf{U}}_{N_S  \times N_S }  + {\bf{Y}}_{N_S  \times N_S } {\bf{Z}}_{N_S  \times N_S }^L )^{ - 1} \nonumber \\ 
                    &\quad {\bf{Y}}_{N_S  \times N_T } ){\bf{V}}_{N_T } \label{eq:Matrix-A11}
\end{align}
Finally, ${\bf{Z}}_{N_R  \times N_R }^L$ is multiplied by both sides of Eq. (\ref{eq:Matrix-A11}), following expressions are obtained. 
Here, ${\bf{V}}_{N_R }$ and ${\bf{H}}^{f}_{N_R  \times N_T }$ are defined as in (\ref{eq:Matrix-A15}) and (\ref{eq:Matrix-A16}), respectively.

\section*{Acknowledgement}
KK and QC are supported by the Ministry of Internal Affairs and Communications of Japan, grant number JPJ000254. 
ST and GG are supported in part by EU H2020 RISE-6G project, grant number 101017011. 
GG is also supported by the Royal Society Industry Fellowship, grant number INF\textbackslash R2\textbackslash192066, 
and by the EPSRC, grant numbers EP/V048937/1 and EP/X038491/1.
%


\end{document}